\documentclass[10pt]{article}

\usepackage{amsmath, amsthm, amsfonts, amssymb}

\newtheorem{thm}{Theorem}[section]

\newcommand{\semifk}{$\left\{P^V_t\right\}_{t\geq 0}$}
\newcommand{\semi}{$\left\{T(t)\right\}_{t\geq 0}\;$}
\newcommand{\semia}{$\left\{T^{*}(t)\right\}_{t\geq 0}\;$}

\title{{\bf $L^\infty$-UNIQUENESS OF GENERALIZED SCHR\"ODINGER OPERATORS}}
\author
{Ludovic Dan LEMLE\thanks{Institut Camille Jordan UMR 5208 (CNRS), Universit\'e
Claude Bernard Lyon1, 69622 Villeurbanne, France and Engineering Faculty of Hunedoara, "Politehnica"
University of Timi\c soara, 331128 Hunedoara, Romania\quad e-mail: {\tt lemle\_dan@yahoo.com}}
}

\date{version 19 December 2007}

\def\D{\cal{D}}
\def\B{\cal{B}}
\def\E{\mathbb{E}}
\def\R{\mathbb{R}}

\begin{document}

\maketitle

\begin{abstract}
\noindent The main purpose of this paper is to show that the generalized Schr\"odinger operator ${\cal A}^Vf=\frac{1}{2}\Delta f+b\nabla f-Vf$, $f\in C_0^\infty(\R^d)$, is a pre-generator for which we can prove its $L^\infty\left(\R^d,dx\right)$-uniqueness. Moreover, we prove the $L^1(\R^d,dx)$-uniqueness of weak solutions for the Fokker-Planck equation associated with this pre-generator.
\end{abstract}
%{\bf 2000 AMS Subject Classification} Primary: 47D03, 47F05. Secondary: 60J60.

\section{Preliminaries}

Let $E$ be a Polish space equipped with a $\sigma$-finite measure
$\mu$ on its Borel $\sigma$-field $\B$. It is well known that, for a $C_0$-semigroup \semi on $L^1(E,d\mu)$, its
adjoint semigroup \semia is  no longer strongly continuous on the
dual topological space $L^\infty(E,d\mu)$ of $L^1(E,d\mu)$ with respect to the
strong dual topology of $L^\infty(E,d\mu)$. In
\cite{wu-zhang'06} {\sc Wu} and {\sc  Zhang} introduce on $L^\infty(E,d\mu)$ {\it the topology of uniform convergence on compact subsets of $(L^1(E,d\mu),\|\:.\:\|_1)$}, denoted by ${\cal
C}(L^\infty,L^1)$,  for which the usual semigroups in the literature
becomes $C_0$-semigroups. If \semi is a $C_0$-semigroup on
$L^1\left(E,\mu\right)$ with generator $\cal L$, then \semia is a
$C_0$-semigroup on $\left(L^\infty(E,d\mu),{\cal
C}\left(L^\infty,L^1\right)\right)$ with generator ${\cal
L}^{*}$. Moreover, on can proove that $\left(L^\infty(E,d\mu),{\cal
C}\left(L^\infty,L^1\right)\right)$ is complete and that the
topological dual of $\left(L^\infty(E,d\mu),{\cal
C}\left(L^\infty,L^1\right)\right)$ is
$\left(L^1(E,d\mu),\|\:.\:\|_1\right)$.\\
Let ${\cal A}:{{\cal D}}\longrightarrow{L^\infty(E,d\mu)}$ be a linear operator with
its domain $\D$ dense in $\left(L^\infty(E,d\mu),{\cal
C}\left(L^\infty,L^1\right)\right)$. $\cal A$ is said to be a {\it
pre-generator} in $\left(L^\infty(E,d\mu),{\cal
C}\left(L^\infty,L^1\right)\right)$, if there exists some $C_0$-semigroup on $\left(L^\infty(E,d\mu),{\cal
C}\left(L^\infty,L^1\right)\right)$ such
that its generator $\cal L$ extends $\cal A$. We say that $\cal A$
is an {\it essential generator} in $\left(L^\infty(E,d\mu),{\cal
C}\left(L^\infty,L^1\right)\right)$ (or $\left(L^\infty(E,d\mu),{\cal
C}\left(L^\infty,L^1\right)\right)$-unique), if $\cal
A$ is closable and its closure $\overline{\cal A}$ with respect to
${\cal C}\left(L^\infty,L^1\right)$ is the generator of some $C_0$-semigroup on $\left(L^\infty(E,d\mu),{\cal
C}\left(L^\infty,L^1\right)\right)$. This
uniqueness notion was studied in {\sc Arendt} \cite{arendt'86}, {\sc
Eberle} \cite{eberle'97}, {\sc Djellout} \cite{djellout'97}, {\sc
R\"ockner} \cite{rockner'98}, {\sc Wu} \cite{wu'98} and
\cite{wu'99} and others in the Banach spaces setting and {\sc Wu} and {\sc Zhang} \cite{wu-zhang'06} and {\sc Lemle} \cite{lemle'07} in the case of locally convex spaces.

\section{$L^\infty$-uniqueness of generalized Schr\"odinger operators}
In this note we consider the generalized Schr\"odinger operator
\begin{equation}
{\cal A}^Vf:=\frac{1}{2}\Delta f+b\nabla f-Vf\quad,\quad\forall f\in C_0^\infty(\R^d)
\end{equation}
where $b:\R^d\rightarrow\R^d$ is a measurable and locally bounded vector field and $V:\R^d\rightarrow\R$ is a locally bounded potential. In the case where $V=0$, the essential self-adjointness of ${\cal A}:=\frac{1}{2}\Delta+b\nabla$ has been completely charaterized in the works of {\sc Wielens} \cite{wielens'85} and {\sc Liskevitch} \cite{liskevitch'99}. $L^1$-uniqueness of this operator has been introduced and studied by {\sc Wu} \cite{wu'99}, its $L^p$-uniqueness has been studied by {\sc Eberle} \cite{eberle'97} and {\sc Djellout} \cite{djellout'97} for $p\in[1,\infty)$ and by {\sc Wu} and {\sc Zhang} \cite{wu-zhang'06} for $p=\infty$.\\
Our purpose is to find some sufficient condition to assure the $L^\infty(\R^d,dx)$-uniqueness of $({\cal A}^V,C_0^\infty(\R^d))$ with respect to the topology ${\cal C}(L^\infty,L^1)$ in the case where $V\geq 0$.\\
At first, we must remark that the generalized Schr\"odinger operator $({\cal A}^V,C_0^\infty(\R^d))$ is a pre-generator on $L^\infty(\R^d,dx)$. Indeed, if we consider the Feynman-Kac semigroup \semifk given by
\begin{equation}
P_t^Vf(x):=\E^x1_{[t<\tau_e]}f(X_t)e^{-\int\limits_0^t\!V(X_s)\:ds}
\end{equation}
where $(X_t)_{0\leq t<\tau_e}$ is the diffusion generated by $\cal A$ and $\tau_e$ is the explosion time, then by \cite[Theorem 1.4]{wu-zhang'06} it follows that \semifk is a $C_0$-semigroup on $L^\infty(\R^d,dx)$ with respect to the topology ${\cal C}(L^\infty,L^1)$. By Ito's formula on can prove that $f$ belongs to the domain of the generator ${\cal L}^V_{(\infty)}$ of $C_0$-semigroup \semifk on $(L^\infty(\R^d,dx),{\cal C}(L^\infty,L^1))$.\\
The main result of this note is
\begin{thm}
Suppose that there is some mesurable locally bounded function $\beta:\R^{+}\rightarrow\R$ such that
\begin{equation}
\frac{b(x)x}{|x|}\geq\beta(|x|)\quad,\quad\forall x\in\R^d,\:x\neq 0.
\end{equation}
Let $\tilde{\beta}(r)=\beta(r)+\frac{d-1}{2r}$. If the one-dimensional diffusion operator
\begin{equation}
{\cal A}^V_1=\frac{1}{2}\frac{d^2}{dr^2}+\tilde{\beta}(r)\frac{d}{dr}-V(r)
\end{equation}
is $L^\infty(0,\infty;dx)$-unique, then $\left({\cal A}^V,C_0^\infty(\R^d)\right)$ is $L^\infty(\R^d,dx)$-unique.\\
Moreover, for any $f\in L^1(\R^d,dx)$ the Fokker-Planck equation
\begin{equation}
\left\lbrace\begin{array}{l}
\partial_tu(t,x)=\frac{1}{2}\Delta u(t,x)-\left(div\:b+V\right)u(t,x)\\
u(0,x)=f(x)
\end{array}
\right.
\end{equation}
has one $L^1(\R^d,dx)$-unique weak solution given by $u(t,x)=P^V_tf(x)$.
\end{thm} 

\bibliographystyle{plain}

\end{document}